\documentclass[anonymous]{llncs}
\usepackage{amsmath,graphicx,multicol,multirow,booktabs}
\usepackage{bbm}
\usepackage[linkcolor=black,anchorcolor=black, citecolor=black]{hyperref}
\usepackage{amssymb}
\usepackage{booktabs} 
\usepackage[table]{xcolor} 
\usepackage{threeparttable} 
\usepackage{enumitem}
\usepackage{adjustbox} 
\begin{document}
\title{Core-Periphery Principle Guided State Space Model for Functional Connectome Classification }
\titlerunning{Core-Periphery Principle Guided State Space Model for Functional Connectome Classification}
%
\author{Minheng Chen \inst{1}  \and Xiaowei Yu\inst{1} \and Jing Zhang \inst{1}\and Tong Chen \inst{1} \and Chao Cao \inst{1}\and Yan Zhuang \inst{1}\and Yanjun Lyu\inst{1}\and  Lu Zhang \inst{2} \and Tianming Liu \inst{3}\and Dajiang Zhu\inst{1}}
%
\authorrunning{ M. Chen et al.}
%
\institute{ Department of Computer Science and Engineering, University of Texas at Arlington, USA \\
\and
Department of Computer Science, Indiana University Indianapolis, USA
\and
School of Computing,  University of Georgia, USA
}
\maketitle              
\begin{abstract}
Understanding the organization of human brain networks has become a central focus in neuroscience, particularly in the study of functional connectivity, which plays a crucial role in diagnosing neurological disorders. Advances in functional magnetic resonance imaging and machine learning techniques have significantly improved brain network analysis. However, traditional machine learning approaches struggle to capture the complex relationships between brain regions, while deep learning methods, particularly Transformer-based models, face computational challenges due to their quadratic complexity in long-sequence modeling.
To address these limitations, we propose a Core-Periphery State-Space Model (CP-SSM), an innovative framework for functional connectome classification. Specifically, we introduce Mamba, a selective state-space model with linear complexity, to effectively capture long-range dependencies in functional brain networks. Furthermore, inspired by the core-periphery (CP) organization, a fundamental characteristic of brain networks that enhances efficient information transmission, we design CP-MoE, a CP-guided Mixture-of-Experts that improves the representation learning of brain connectivity patterns.
We evaluate CP-SSM on two benchmark fMRI datasets: ABIDE and ADNI. Experimental results demonstrate that CP-SSM surpasses Transformer-based models in classification performance while significantly reducing computational complexity. These findings highlight the effectiveness and efficiency of CP-SSM in modeling brain functional connectivity, offering a promising direction for neuroimaging-based neurological disease diagnosis.
\end{abstract}

\section{Introduction}
\label{sec:intro}
The vast assemblage of neurons in human brain forms a complex, interconnected network that achieves a remarkable balance between regional specialization and global functional integration, enabling diverse cognitive and behavioral processes~\cite{fornito2013graph,fornito2015connectomics,stevens1979neuron}.
Understanding the organization of these neural networks has become a central focus of contemporary neuroscience.
By studying and analyzing brain networks, neuroscientists can gain deeper insights into the structural and functional architecture of the human brain, as well as how network dynamics influence the onset, expression, and manifestation of neurological diseases~\cite{chen2025usingstructuralsimilaritykolmogorovarnold,van2019cross}.
The recent advancements in neuroimaging techniques have enabled detailed mapping and analysis of brain connections, providing unprecedented insights into the intricate architecture of brain connections. Among these imaging modalities, functional magnetic resonance imaging (fMRI), which uses blood oxygen level-dependent (BOLD) signals to non-invasively assess the intrinsic functional connectivity of brain, is one of the most commonly used modalities for brain network analysis~\cite{bandettini2012twenty,glover2011overview}.
In recent years, numerous studies have focused on utilizing functional connectivity (FC)—calculated as the correlation of BOLD signals across different brain regions—for the diagnosis of neurological diseases and the identification of potential biomarkers~\cite{hohenfeld2018resting}.

Machine learning methods, \textit{e.g.} support vector machines and random forests, are widely employed for classifying functional connectivity patterns. However, traditional machine learning approaches face inherent limitations in capturing the complex and intricate relationships between different brain regions. 
In contrast, deep learning-based methods leverage specifically designed neural network architectures to computationally model brain functional connectivity~\cite{kawahara2017brainnetcnn,kim2020understanding,zhang2025brain,zhou2024multi}. By harnessing the powerful representational capabilities of deep neural networks, these methods often achieve considerable success in the analysis of brain functional connectomes.
The Transformer model architecture, initially developed for natural language processing, has garnered significant attention in the field of medical imaging, due to its capacity to capture long-range dependencies. 
Recent studies have demonstrated that Transformer-based methods substantially outperform other learning-based benchmarks in classifying neurological disorders~\cite{kan2022brain,kong2024multi,peng2024gbt,zhang2025classification}. 

Although existing methods can effectively model functional connectivity patterns of the human brain and achieve competitive results in diagnosing neurological diseases such as Alzheimer's disease (AD) and autism spectrum disorder(ASD), they still have the following shortcomings:
First, existing works rely heavily on increasingly complex network architectures, which not only lead to overfitting problems due to inductive bias, but also hinder their application on long sequences by computational complexity such as the quadratic computational cost associated with the attention mechanism. 
These limitations have driven the pursuit of more computationally efficient alternatives to the Transformer architecture—methods that not only offer improved computational complexity but also excel at capturing long-range dependencies and preserving strong representation learning capabilities.
Second, current network models often fail to incorporate the inherent characteristics of brain function during computational modeling, leading to suboptimal performance in brain network analysis. This oversight highlights the need for approaches that better align with the unique functional and structural attributes of the brain.

In this paper, we address the aforementioned limitations by introducing a core-periphery principle guided state-space model (CP-SSM) for functional connectome classification.
Specifically, 1) we introduce a highly promising long-sequence modeling method with linear complexity based on a selective state space model named Mamba~\cite{gu2023mamba} to capture the functional connectivity relationships between different brain regions and  identify variations in functional brain patterns across individuals.
2) Core-periphery (CP) organization is a ubiquitous feature of the brain functional network in humans and other mammals. 
It has been widely demonstrated to enhance the efficiency of information transmission and communication in biological integration processes~\cite{csermely2013structure}. 
Drawing inspiration from this principle, we propose CP-MoE, which leverages CP organization to guide the redesign of Mixture-of-Experts (MoE) model, thereby effectively enhancing the representation capabilities of the networks.
3) Experimental results on the ABIDE and ADNI datasets demonstrate that the proposed method surpasses existing Transformer-based approaches while achieving lower computational complexity, highlighting its effectiveness and efficiency.
\begin{figure}[htb]

  \centering
  \centerline{\includegraphics[width=\linewidth]{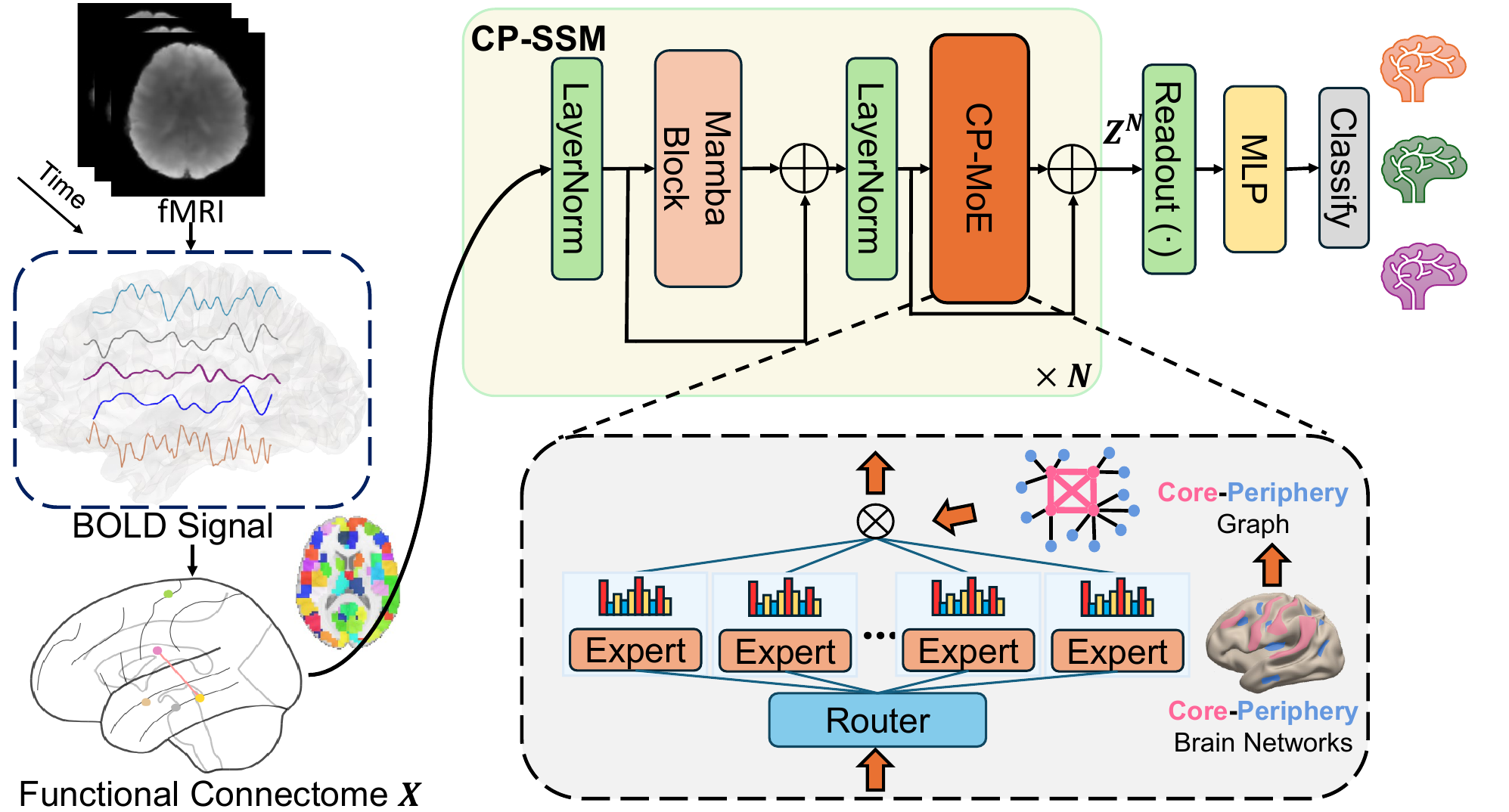}}
%
\caption{Overall architecture of the proposed method. 
Our approach is founded on a state-space model, with its key components comprising a Mamba block and a Core-Periphery principle guided MoE.}
\label{fig:overall}
\end{figure}
\vspace{-0.8cm}
\section{Material and Method}
The overall pipeline of the proposed method is shown in Fig.~\ref{fig:overall}.
The FC matrix $X\in \mathbb{R}^{V\times V}$($V$ is the number of ROIs), derived by calculating the Pearson cross-correlation of the preprocessed BOLD signals across different brain regions, is first processed through $N$ CP-SSM blocks to generate the node-level embedding representation $Z\in \mathbb{R}^{V\times V}$. Each CP-SSM block comprises two key components: an SSM module and a CP-MoE block. Subsequently, the output is passed through an Orthonormal Clustering Readout function~\cite{kan2022brain} to obtain the graph-level embedding. The embedding is then fed into a fully connected layer, with the final probability predictions generated through a softmax layer. The model is trained using cross-entropy loss.
In the following sections, we will begin by detailing our data acquisition and preprocessing pipelines. Next, we will outline the framework of the proposed CP-SSM. Finally, we will provide an in-depth description of the architectural details of CP-MoE.
\subsection{Data Acquisition and Preprocessing}
In this study, we evaluated the proposed method on two fMRI datasets, assessing its performance in diagnosing two neurological conditions: autism spectrum disorder (ASD) and mild cognitive impairment (MCI)—the prodromal stage of AD.
(a) \textit{Autism Brain Imaging Data Exchange} (ABIDE)~\cite{craddock2013neuro}: This dataset collects resting-state fMRI data from 17 international sites. Based on the given quality control scores, 1009 subjects (516 with ASD and 493 with NC) from 1,112 subjects were selected.
The dataset, preprocessed by the Configurable Pipeline for the Analysis of Connectomes (CPAC) tool, underwent band-pass filtering (0.01 - 0.1Hz) without global signal regression. The brain was parcellated using the Craddock 200 atlas~\cite{craddock2012whole}.
(b) \textit{Alzheimer’s Disease Neuroimaging Initiative}(ADNI)~\cite{mueller2005alzheimer}: 440 subjects(215 MCI, 225 NC) were selected based on quality control.
Each subject’s data underwent the same standard preprocessing procedures as detailed in~\cite{zhu2014connectome}
The Destrieux Atlas was then applied for parcellation.
Subject demographics and preprocessing details of the ADNI dataset can be found in the \textbf{supplementary material}.

\subsection{State Space Model for Functional Connectome Classification}
A State Space Model (SSM) is commonly defined as a linear time-invariant system that maps a one-dimensional input sequence $x(t)\in \mathbb{R}$ to an output response $y(t)\in \mathbb{R}$ through an underlying latent state representation $h(t)\in \mathbb{R}^M$, as formalized in Eq.~\ref{eq::ssm}, where $A \in R^{M\times M}$ , $B \in \mathbb{R}^{M\times1}$, $C \in \mathbb{R}^{1\times M}$ , and $D \in \mathbb{R}^1$ are the weighting parameters.
\begin{equation}
\label{eq::ssm}
\begin{aligned}
    h'(t) &= \mathbf{A} h(t) + \mathbf{B} u(t), \\
    y(t) &= \mathbf{C} h(t) + \mathbf{D} u(t),
\end{aligned}
\end{equation}
Mamba~\cite{gu2023mamba}, a recently proposed variant of SSM, exhibits a remarkable capacity for processing long sequential data with linear computational complexity, making it highly efficient in modeling local interactions between adjacent nodes. In this study, we propose utilizing Mamba modules to characterize brain network connectivity patterns, enabling the capture of long-range dependencies across brain regions that would otherwise be computationally prohibitive to model using self-attention mechanisms.
The input functional connectivity matrix X first undergoes Layer Normalization, producing $X_n$. Subsequently, $X_n$ is processed through a Mamba block, which consists of a linear layer, a SiLU~\cite{hendrycks2016gaussian} activation function, a one dimensional convolution layer, and an SSM layer~\cite{Chen2024mxt}. The output $X'\in \mathbb{R}^{V\times V}$ is then combined with $X_n$ via element-wise addition to generate the latent representation $H\in \mathbb{R}^{V\times V}$. This entire process can be mathematically formulated as follows:
\begin{equation}
\begin{aligned}
        X'&=SSM(conv(SiLU(linear(X_n)))\\
        H&=X'+X_n
\end{aligned}
\end{equation}
In our architecture, the Feedforward Network (FFN) commonly used in Transformers is replaced with the proposed CP-MoE, a widely adopted approach in the field of Sparse MoE (SMoE)~\cite{yun2024flexmoe}.The details of CP-MoE will be introduced later.
After normalization, the latent representation $H$ is transformed into $H_n$. Next, $H_n$ is processed through CP-MoE, yielding $H'$. The output $H'$is then combined with $H_n$ via element-wise addition to generate the node-level embedding Z. Since our model consists of $N$ cascaded CP-SSM, this process is iteratively repeated $N$ times to obtain the final, more expressive node feature $Z^N \in \mathbb{R}^{V \times V}$.
\subsection{Core-periphery Principle Guided Mixture-of-Experts }
Mixture-of-Experts has advanced by integrating sparsity into its architecture, enhancing both computational efficiency and overall performance. 
SMoE improves scalability by selectively activating only the most relevant experts for a given task, thereby minimizing computational overhead. This approach is especially advantageous for processing complex, high-dimensional datasets across various applications.
Core-periphery (CP) organization is a fundamental structural characteristic of functional brain networks in humans and other mammals. In this framework, nodes within the brain network are categorized into core nodes and peripheral nodes. Core nodes exhibit dense interconnectivity, facilitating efficient information processing, whereas peripheral nodes maintain sparser connections, both among themselves and with core nodes.
Recent studies~\cite{yu2024gyri} suggest that the brain’s gyri and sulci function collaboratively within a core-periphery network, further reinforcing the role of CP organization in neural communication. Empirical evidence has demonstrated that CP architecture significantly enhances information transfer efficiency and biological integration processing~\cite{csermely2013structure}.
Motivated by this principle, we aim to redesign the expert selection mechanism in MoE under the guidance of CP principle. Specifically, we propose leveraging the CP framework to guide the router in expert assignment, ensuring a more structured and biologically inspired approach to model optimization.

First, we construct a CP graph $G$ to facilitate expert selection based on the CP principle. To achieve this, we introduce a parameter core node rate $r$, which partitions the graph's nodes into two distinct sets: core nodes $\mathcal{C}$ and peripheral nodes $\mathcal{P}$. Notably, when $r=1$, the resulting CP graph becomes fully connected. Consistent with the definition in~\cite{yu2024core}, $G$ can be represented as:
\[
G(i,j) =
\begin{cases}
1, & \text{if } (i, j) \in \mathcal{C} \times \mathcal{C}\text{ or } (i, j) \in \mathcal{C} \times \mathcal{P} \text{ or } (i, j) \in \mathcal{P} \times \mathcal{C} \\
0, & \text{if } (i, j) \in \mathcal{P} \times \mathcal{P}
\end{cases}
\]
Thus, the Top-$k$ expert selection of the router $\mathcal{R}(\cdot)$ in MoE, determined by the highest scores from softmax function with the learnable gating function $g(\cdot)$, can be formally expressed as follows:
\begin{equation}
     \mathbf{y} = G\cdot\sum_{i=1}^{E} \mathcal{R}(\mathbf{x})_i \cdot f_i(\mathbf{x})
\end{equation}
\begin{equation}
    \mathcal{R}(\mathbf{x}) = \text{Top-}k(\text{softmax}(g(\mathbf{x})), k)
\end{equation}
\begin{equation}
    \text{Top-}k(\mathbf{v}, k) = 
    \begin{cases} 
        \mathbf{v}, & \text{if } \mathbf{v} \text{ is in the top } k, \\
        0, & \text{otherwise}.
    \end{cases}
\end{equation}
where where $\mathbf{x}$ and $\mathbf{y}$ denote the input and output of MoE, respectively. $E$ represents the total number of experts, and $i$ denotes the index of a specific expert.
\section{Experiments}
\subsection{Experiment Settings}
We partition each dataset into 70\% for training, 10\% for validation, and 20\% for testing. 
For evaluation on the test set, we select the epoch that achieves the highest AUROC score on the validation set.
We evaluate the performance of our proposed method against several baseline approaches, including three traditional machine learning methods: support vector machine (SVM), random forest (RF), and XGBoost; two CNN/GNN-based methods: FBNETGEN~\cite{kan2022fbnetgen} and BrainNetCNN~\cite{kawahara2017brainnetcnn}; and four Transformer-based methods: VanillaTF~\cite{kan2022brain}, BrainNetTF~\cite{kan2022brain}, Com-BrainTF~\cite{bannadabhavi2023community}, and GBT~\cite{peng2024gbt}.
Details regarding the hyperparameter settings and implementation of these baseline methods are provided in the supplementary material.

\noindent\textbf{Implementation details.}
As for our method, we set the state-space dimension of the Mamba block to 16, the expansion factor to 2, the number of CP-SSM block $N=2$, and the convolution kernel dimension to 4. 
The CP-MoE consists of 8 experts, employing a Top-2 selection strategy.
For dataset-specific configurations, we set the core node rate to 0.2 on the ABIDE dataset. 
However, due to the inherent sparsity of FC in ADNI dataset (an illustraion can be found in supplementary material), we increase the core node rate to 0.8 to better accommodate its characteristics.
The training of CP-SSM is performed using the Adam optimizer, initialized with a learning rate of 10e-4 and a weight decay of 10e-4. A cosine annealing schedule is employed for learning rate adjustment, starting at 10e-4 and progressively reducing to 10e-5 without a warm-up phase. The training process utilizes a batch size of 64 and spans 200 epochs.
All experiments were conducted on a PC equipped with an NVIDIA RTX 6000 Ada GPU and a 3.6-GHz Intel Core i7 processor.



\subsection{Performance Comparison with Baseline Methods}
\begin{table}[h!]
\scriptsize

    \centering
    \caption{Performance comparison with different baselines on ADNI and ABIDE.}
    \begin{tabular}[width=\linewidth]{c|c|c|c|c|c|c|c|c}
      \hline
     \multicolumn{1}{c|}{\multirow{2}{*}{Methods}} &  \multicolumn{4}{|c}{\textbf{Dataset: ABIDE}} & \multicolumn{4}{|c}{\textbf{Dataset: ADNI}} \\
        \cline{2-9}&AUROC& ACC&SEN & SPE  &AUROC& ACC &SEN & SPE \\
    \hline
        SVM   & 70.4$\pm$5.2 & 63.3$\pm$5.2& 64.8$\pm$7.1 &61.6$\pm$7.0&65.1$\pm$8.2&61.5$\pm$4.4&51.2$\pm$7.9&69.7$\pm$8.2\\
        RF   & 69.2$\pm$4.3 & 63.8$\pm$3.4& 71.0$\pm$5.2 &56.1$\pm$5.2 &67.9$\pm$3.8&63.9$\pm$1.2&55.9$\pm$4.7&71.4$\pm$4.5\\
        XGBoost &71.2$\pm$4.4 &63.4$\pm$5.1& 68.6$\pm$4.9&57.8$\pm$9.1 &65.4$\pm$5.0 &62.7$\pm$1.1 &61.5$\pm$6.0 &63.8$\pm$6.2  \\
      \hline
      FBNETGNN   & 72.9$\pm$5.1 & 65.7$\pm$5.6 & 64.3$\pm$10.6&66.6$\pm$8.2&69.1$\pm$7.9&66.3$\pm$3.9&66.7$\pm$8.1&65.7$\pm$8.4\\
      BrainNetCNN  & 73.2$\pm$3.0 & 66.6$\pm$4.0 & 64.6$\pm$6.2&68.7$\pm$4.8& 65.8$\pm$1.0&65.4$\pm$5.2&60.7$\pm$1.3&68.7$\pm$4.6\\
      \hline
      VanillaTF  & 79.6$\pm$4.6 & 69.8$\pm$6.0 & 64.1$\pm$8.1&\textbf{76.4$\pm$9.1}&73.1$\pm$6.4&69.3$\pm$3.8&68.7$\pm$8.1&70.3$\pm$8.8\\
      BrainNetTF & 79.1$\pm$4.8 & 70.1$\pm$4.9 & 67.9$\pm$5.0&72.2$\pm$6.6& 73.0$\pm$7.4&70.3$\pm$4.7&69.7$\pm$7.6&69.9$\pm$7.7\\
      Com-BrainTF  & 77.3$\pm$4.1& 71.6$\pm$4.5 & 75.1$\pm$11.9&67.4$\pm$9.3&-&-&-&-\\
      GBT  & 78.3$\pm$4.1 & 71.5$\pm$5.8 & 75.5$\pm$14.7 &68.2$\pm$12.8&74.5$\pm$5.5 &71.1$\pm$3.2&69.0$\pm$7.8&73.6$\pm$5.1\\ 
      \rowcolor{gray!20}
      \hline
      
      CP-SSM  &  \textbf{82.4$\pm$2.2}& \textbf{76.9$\pm$1.4} & \textbf{79.3$\pm$8.8}&74.5$\pm$8.4&  \textbf{80.7$\pm$2.6}&\textbf{74.9$\pm$2.6}&\textbf{75.4$\pm$5.7}&\textbf{74.2$\pm$2.6}\\ 
      
      \hline
    \end{tabular}

    \label{tab:table1}
\end{table}
We compare the proposed CP-SSM with baseline methods on the ASD classification task using the ABIDE dataset and on the MCI classification task using the ADNI dataset, respectively.
We repeat the above experimental process ten times on each dataset and report the disease classification results. 
It is important to note that the brain network community division in Com-TF follows the assignments of the Yeo 7-network template~\cite{yeo2011organization}, which is derived from brain network analysis in young individuals. As a result, it is well-suited for the ABIDE dataset (ages 7–64 years, with a median age of 14.7 years across groups). However, our ADNI dataset consists of elderly participants (ages 50.5–82.8 years), making this template less appropriate. Consequently, we did not report Com-TF results on the ADNI dataset.
The results in Table~\ref{tab:table1} demonstrate that CP-SSM consistently outperforms all baselines on both ABIDE and ADNI, 
highlighting its superior capability in modeling functional connectivity for neurological disease diagnosis compared to both traditional machine learning and deep learning-based approaches.

 
\subsection{Ablation Study}
We perform ablation studies to assess the effectiveness of each component in the design of the proposed CP-SSM, with the corresponding experimental results presented in Table 2.
1) w/o CP: we remove the CP mask in CP-MoE, and now the structure of MoE is equivalent to other SMoE methods~\cite{shazeer2017outrageously,yun2024flexmoe};
2) w/o MoE: directly replace all MoE layers in the network with MLPs with the same number of neurons;
3) w/o CP-MoE: CP-MoE is replaced by the FNN in the common Transformer architecture;
4) w/o SSM: replace SSM with the same transformer block as implemented in~\cite{kan2022brain};
5) learnable mask: substitute the core-periphery guided mask with a weighted learnable mask.
Furthermore, replacing the SSM module with a Transformer block led to an increase in the number of model parameters from 11.6M to 12.2M, while classification accuracy decreased by 2\%. This finding highlights the computational efficiency of the SSM module in maintaining model performance while reducing parameter complexity.
\begin{table}[htbp]
\centering
\setlength{\tabcolsep}{10pt} 
\caption{Ablation study of CP-SSM on ABIDE with the best and second-best values in \textbf{boldface} and \underline{underline}, respectively.}
        \begin{tabular}{lcccc}
        \toprule
        \textbf{} & \textbf{AUC} & \textbf{ACC}& \textbf{SEN} &\textbf{SPE} \\
        \midrule
 
        \rowcolor[gray]{0.9}
        CP-SSM & \textbf{82.4$\pm$2.2} & \textbf{76.9$\pm$1.4} & \underline{79.3$\pm$8.8} &\textbf{74.5$\pm$8.4}\\
        \hline
        w/o CP & 79.9$\pm$3.3 & \underline{75.9$\pm$ 2.3}& \textbf{81.9$\pm$7.3}&69.2$\pm$8.9\\
        w/o MoE& \underline{81.1$\pm$3.1}& 74.1$\pm$1.6 & 75.7$\pm$5.7 & 72.5$\pm$8.0\\
        w/o CP-MoE & 80.1$\pm$2.3&73.0$\pm$1.3 & 72.4$\pm$4.5 &\underline{73.0$\pm$1.3}\\
        w/o SSM & \underline{81.1$\pm$2.2}&74.7$\pm$0.8 & 77.3$\pm$5.0 &72.3$\pm$5.3\\
        learnable mask & 79.4$\pm$2.3 & 74.0$\pm$1.7 & 73.5$\pm$6.2&\textbf{74.5$\pm$4.9} \\
        \bottomrule
        \end{tabular}
        \label{table::ablation}
\end{table}

\begin{figure}[htbp]
    \centering
    \begin{minipage}[t]{0.45\textwidth} 
        \centering
        \includegraphics[width=\textwidth]{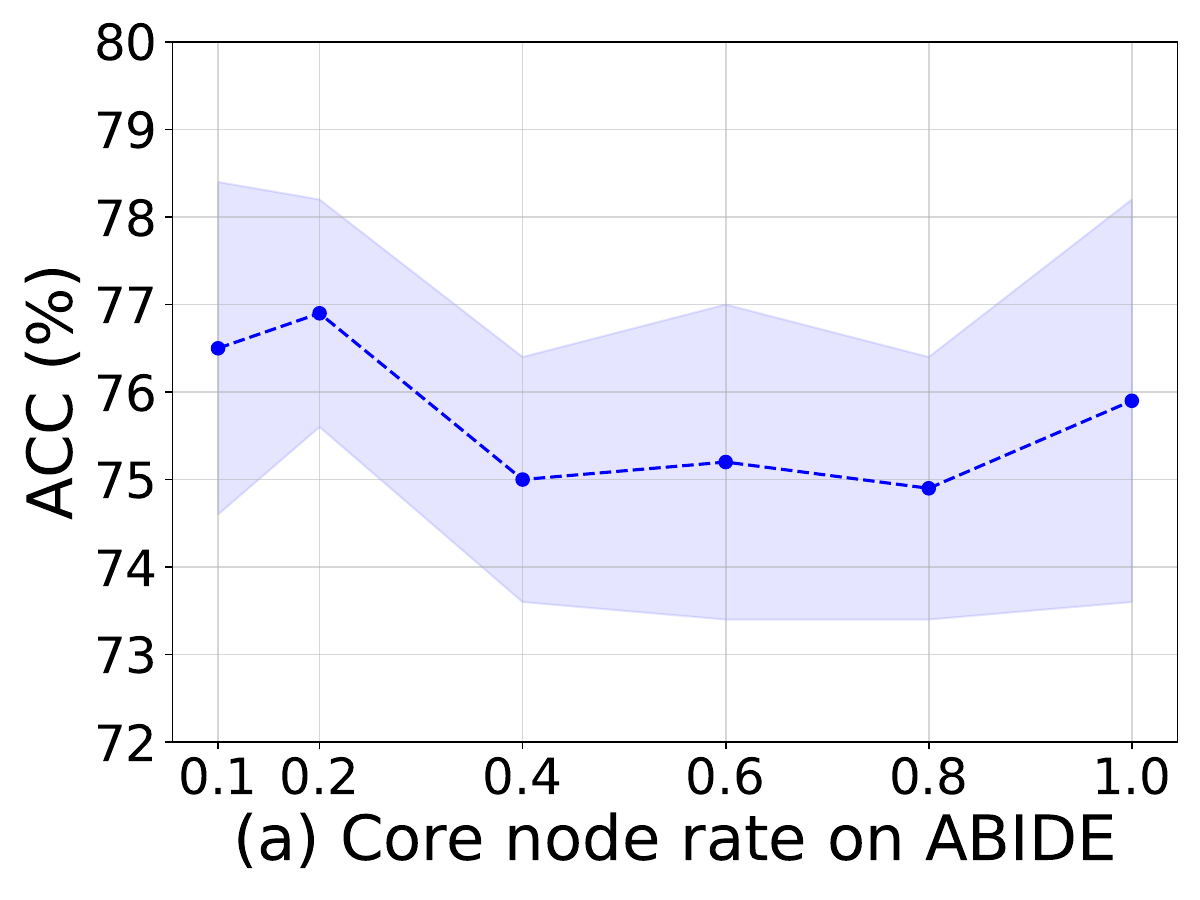} 
        \label{fig:left}
    \end{minipage}%
    \hfill 
    \begin{minipage}[t]{0.45\textwidth} 
        \centering
        \includegraphics[width=\textwidth]{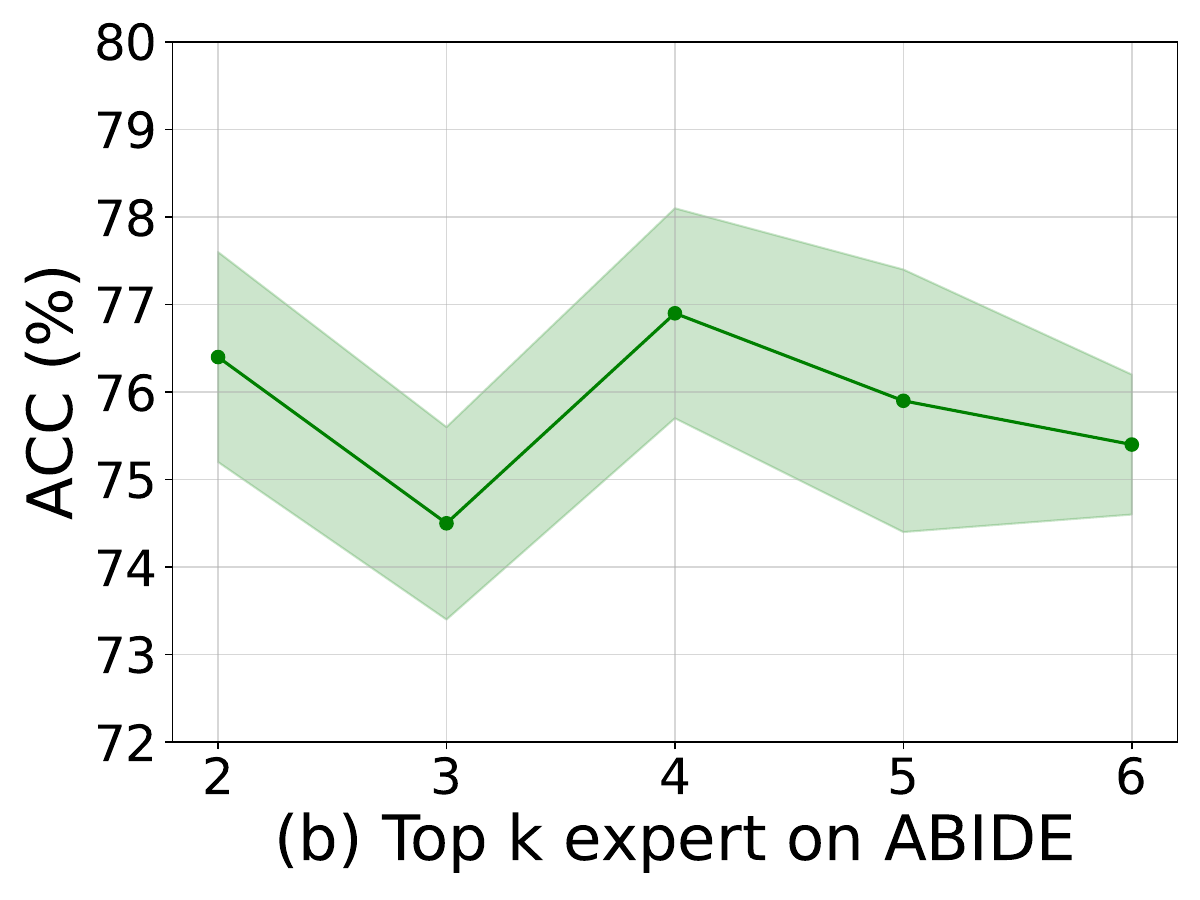} 
        \label{fig:right}
    \end{minipage}
    \vspace{-0.7cm}
            \caption{Sensitivity analysis of CP-SSM on ABIDE. The hyperparameters include core node rate $r$ and top-$k$ expert selection.}
\label{fig::sensitivity}
\end{figure}

\noindent\textbf{Sensitivity analysis.}
The sensitivity analysis of the proposed method on ABIDE is shown in Fig.~\ref{fig::sensitivity}.
We examined the core node rate $r$ and top-$k$ selection.
We find that a) using more core nodes does not always guarantee higher performance than the increase in complexity, which suggests that using a more moderate core node rate is a suitable choice for model performance gain. b) Consistent with the results of many other studies, the top-2 gating network is the most effective.
\vspace{-0.6cm}
\vspace{-0.8cm}
\subsection{Neuroscientific Analysis}
\begin{figure}[htb]

  \centering
  \centerline{\includegraphics[width=\linewidth]{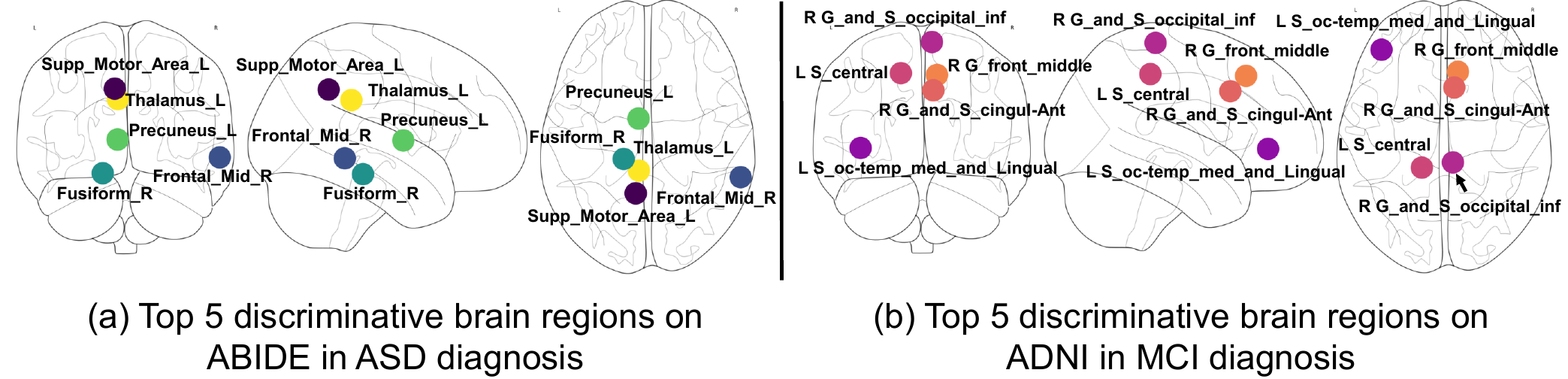}}
%
\caption{Top 5 discriminative brain regions derived from the learnable weight in the last CP-SSM layer on (a) ABIDE and (b) ADNI, with different colormap intensities reflecting relative significance.}
\label{fig:roi}
\end{figure}

As shown in Fig~\ref{fig:roi}, we visualize the learnable weights of the last layer of CP-SSM block to show the top 5 rated brain regions for ASD diagnosis and MCI diagnosis, respectively. 
It is important to note that the Craddock 200 atlas used for the ABIDE dataset is constructed using the normalized cut spectral clustering algorithm, meaning that its clusters do not have precise anatomical labels. In the figure, we present the brain region names from the Automated Anatomical Labeling (AAL) atlas that exhibit the highest overlap with each corresponding region.
\begin{itemize}
    \item \textbf{ASD analysis}: the identified regions—Supplementary Motor Area (SMA), Thalamus, Precuneus, Fusiform Gyrus, and Middle Frontal Gyrus—are associated with social cognition and motor functions. Research indicates reduced connectivity in the Precuneus and Middle Frontal Gyrus in individuals with ASD, affecting social behavior and theory of mind~\cite{cheng2015autism}.
    \item \textbf{MCI Analysis}: in the context of Mild Cognitive Impairment (MCI), the highlighted regions—Inferior Occipital Gyrus, Middle Frontal Gyrus, Anterior Cingulate Gyrus, Central Sulcus, and Occipito-temporal \& Lingual Gyrus—exhibit strong associations with neurodegenerative processes and cognitive decline. 
    The findings of the discrimination region are consistent with existing studies~\cite{palejwala2020anatomy,peng2016clinical}.
\end{itemize}
These findings corroborate existing research, highlighting the involvement of these brain regions in ASD and MCI.
\section{Conclusion}
In this paper, we propose a core-periphery principle guided core-periphery principle guided state-space model for functional connectome classification. 
We utilize the Mamba block to model long-range dependencies in brain connectivity and incorporate the core-periphery principle to guide the MoE. To the best of our knowledge, this is the first application of Mamba and MoE in functional brain network analysis.
Extensive experimental results demonstrate that our method not only outperforms other deep learning approaches but also offers strong interpretability, making it a robust and insightful framework for functional brain network analysis.
Future work will focus on validating the effectiveness of our method across a broader range of datasets and assessing its performance on tasks beyond classification.
%
%
%
\bibliographystyle{splncs04}
%
\bibliography{mybib}
\newpage
\appendix
\section{Subject Demographics and Preprocessing details of ADNI}
\begin{table}[htbp]
\centering
\setlength{\tabcolsep}{12pt}
\caption{Subject demographics for ADNI dataset}
\begin{threeparttable}
\begin{tabular}{lcc}
\toprule
\multirow{2}{*}{}  & \multicolumn{2}{c}{Mean $\pm$ standard deviation}\\
\cmidrule(r){2-3} 
 & Healthy controls & Patients with MCI  \\
\midrule
Sample size & 225 & 215  \\
Male/female & 111/114 & 115/100  \\

\midrule
Age (years) & 71.76 $\pm$ 6.39 & 71.34 $\pm$ 7.68  \\
Male age (years)& 72.87 $\pm$ 6.44 & 72.27 $\pm$ 7.14  \\
Female age (years) & 70.62$\pm$ 6.14&  70.27 $\pm$8.12 \\
\bottomrule
\end{tabular}
\end{threeparttable}
\end{table}
The ADNI dataset includes two image modalities: T1-weighted MRI and rs-fMRI. 
we applied skull removal for all modalities. 
For rs-fMRI images, we applied spatial smoothing, slice time correction, temporal pre-whitening, global drift removal and band pass filtering (0.01-0.1 Hz).
The rs-fMRI images were sequentially processed as follows:
(1) converting the DICOM file to NIFTI format data, (2) removing the first 10 time points to exclude data instability due to scanner noise, (3) correcting
slice timing for head motion, (6) spatial smoothing, (7) temporal pre-whitening, (8) regressing out nuisance signals, (9) temporal band passing (0.01–0.1 Hz) to minimize low frequency drift and filter the high frequency noise.
All of these preprocessing steps are implemented using FMRIB Software Library (FSL) FEAT. 
For T1-weighted images, we registered them to same space by FSL FLIRT and then conducted parcellation using FreeSurfer package. 
After the parcellation, we adopted the Destrieux Atlas for ROI labeling.
\section{Implementation Details of the Baselines}
The hyperparameter settings of the three traditional machine learning baseline methods (SVM, RF, XGBoost) are obtained through rigorous grid search. For the ABIDE dataset, the support vector machine uses the RBF kernel, while the linear kernel is used for the ADNI dataset.
For the ABIDE dataset, the Random Forest model is configured with the following hyperparameters: a maximum tree depth of 10, a minimum of 4 samples required per leaf node, a minimum of 2 samples required to split a node, and the number of estimators (trees) set to 200. The XGBoost model is configured with the following best hyperparameters: a learning rate of 0.1, a maximum tree depth of 300 estimators (trees), and a subsample ratio of 0.8.
For the ADNI dataset, the hyperparameters differ slightly: the maximum tree depth remains 10, but the minimum samples per leaf node is reduced to 1, the minimum samples required to split a node is increased to 15, and the number of estimators is set to 300.
The XGBoost model uses a more detailed configuration: a learning rate of 0.1, a maximum tree depth of 3, 200 estimators, a subsample ratio of 1.0, a column sampling ratio by tree (colsample\_bytree) of 0.9, a gamma value of 0.1, a minimum child weight of 3, and regularization terms reg\_alpha set to 0 and reg\_lambda set to 1.
Other deep learning-based baseline methods all use the code from their official repositories\footnote{https://github.com/Wayfear/BrainNetworkTransformer}\footnote{https://github.com/ubc-tea/Com-BrainTF}\footnote{https://github.com/CUHK-AIM-Group/GBT}, and the settings of hyperparameters also follow the original papers.

\begin{figure}[htbp]
    \centering
    \begin{minipage}[t]{0.5\textwidth} 
        \centering
        \includegraphics[width=\textwidth]{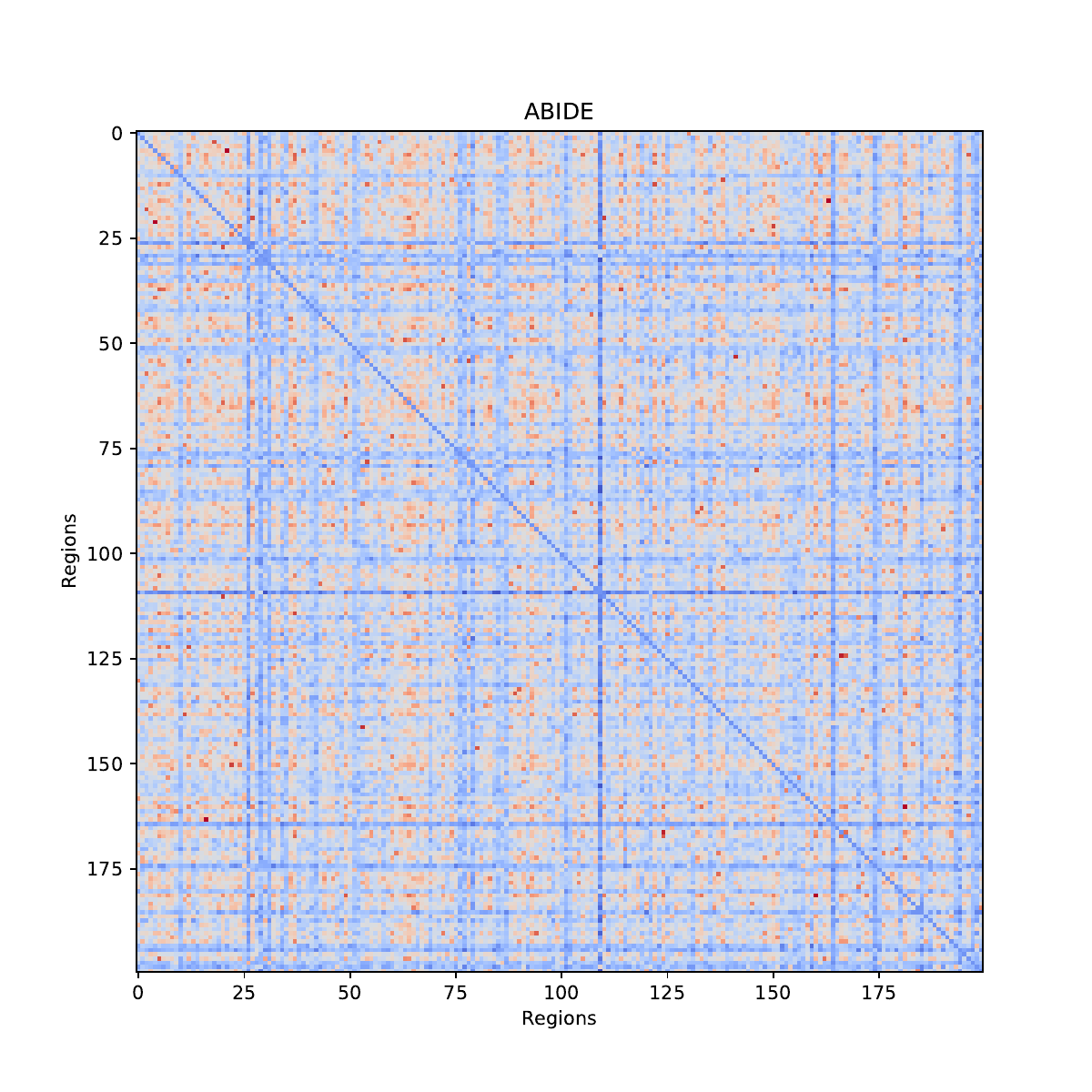} 
        \label{fig:abide}
    \end{minipage}%
    \hfill 
    \begin{minipage}[t]{0.5\textwidth} 
        \centering
        \includegraphics[width=\textwidth]{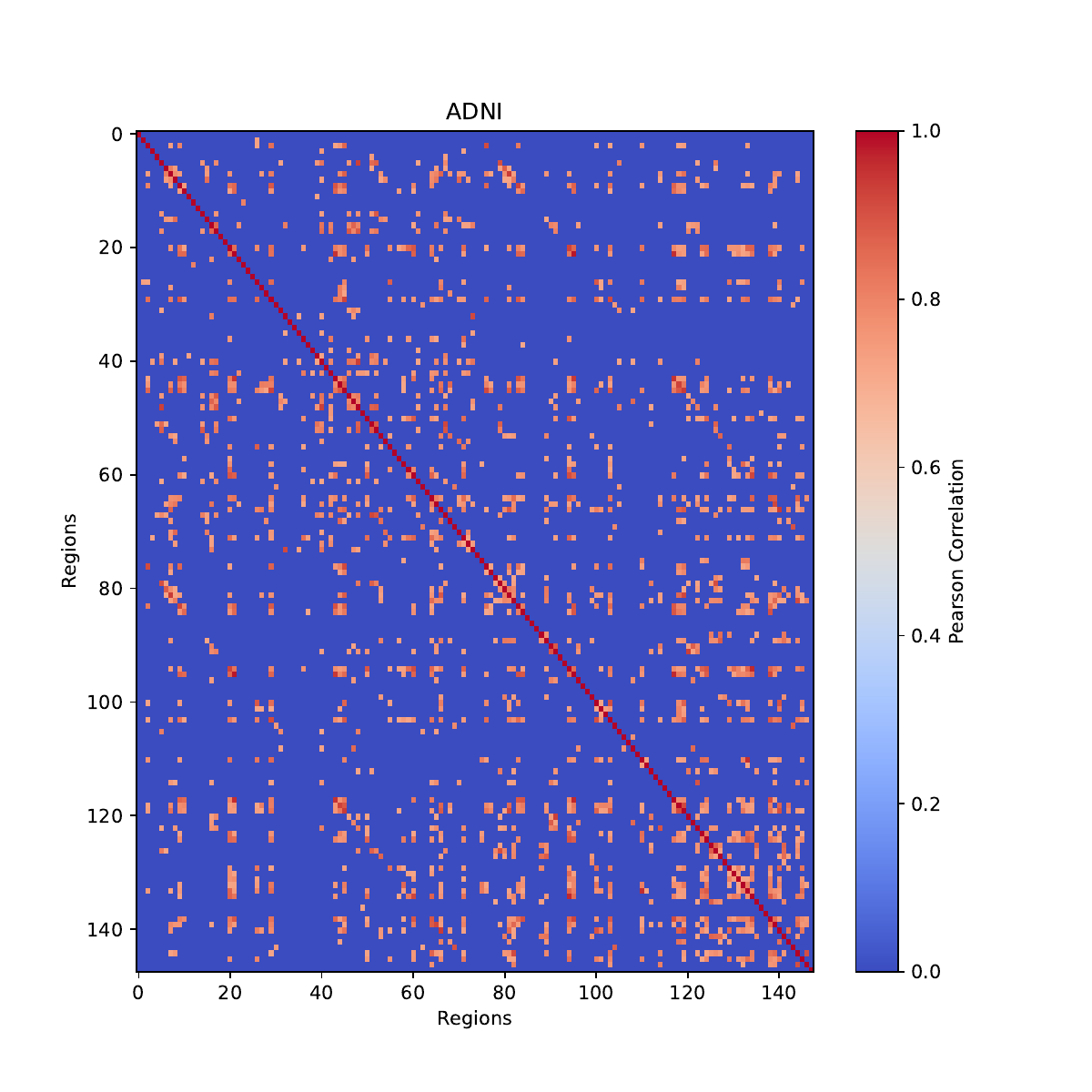} 
        \label{fig:adni}
    \end{minipage}
    \vspace{-0.7cm}
            \caption{Visualization of functional connectivity for a randomly selected subject from the ABIDE and ADNI datasets respectively.}
\label{fig::fc}
\end{figure}

As for our method, we set the state-space dimension of the Mamba block to 16, the expansion factor to 2, and the convolution kernel dimension to 4. The CP-MoE consists of 8 experts, employing a Top-2 selection strategy.
For dataset-specific configurations, we set the core node rate to 0.2 on the ABIDE dataset. However, due to the inherent sparsity of FC in ADNI dataset (Fig.~\ref{fig::fc}), we increase the core node rate to 0.8 to better accommodate its characteristics.
\end{document}